# Feline coronaviruses associated with feline infectious peritonitis have modifications to spike protein activation sites at two discrete positions

**Beth N. Licitra[1], Kelly L. Sams[1], Donald W. Lee[2] and Gary R.Whittaker[1*]**

[1] Department of Microbiology & Immunology, College of Veterinary Medicine, Cornell University, Ithaca, NY, USA.

[2] School of Chemical & Biomolecular Engineering, Cornell University, Ithaca, NY, USA.

* corresponding author. C4 127 Veterinary Medical Center, 930 Campus Rd, Ithaca NY 14853; grw7@cornell.edu

**Abstract.** Feline infectious peritonitis (FIP) is associated with mutations in the feline coronavirus (FCoV) genome that are thought to convert the subclinical feline enteric coronavirus (FECV) into the lethal feline infectious peritonitis virus (FIPV). A key feature of FIPV, not shared with FECV, is the productive infection of macrophages. Therefore mutations in proteins that govern cell tropism, such as the spike glycoprotein, may play an important role in FIP progression. In a prior study, involving a limited number of samples, we have shown an association of FIP with mutations in the protease cleavage-activation site located between the receptor-binding and fusion domains of the FCoV spike (S1/S2). Here, we extend these studies to investigate a larger sample set and to obtain a more refined analysis of the mutations at this S1/S2 site. Our larger data set more clearly shows that the mutations acquired by FIPV at S1/S2 are also accompanied by additional mutations at a second protease cleavage-activation site located in the fusion domain (S2'), adjacent to the viral fusion peptide. Overall, our data indicate a pattern of mutations across the two protease recognition sites that results in substitutions, and/or altered recognition, of critical basic/polar amino acid residues needed for virus activation in the enteric tract. Typically, FIPVs have substitutions of non-polar, aliphatic or aromatic residues in the protease recognition sites. These changes likely modulate the proteolytic activation of the virus and its ability to productively infect macrophages *in vivo*.

## Introduction

Feline infectious peritonitis (FIP) is a fatal infection of cats caused by a feline coronavirus (FCoV) (11). FCoV infections are common, especially in high-density housing situations such as animal shelters and breeding facilities. There are two biotypes of FCoV, classified as either feline enteric coronavirus (FECV) or feline infectious peritonitis virus (FIPV). The biotypes differ based on the severity of infection in cats. Typically, the FECV biotype of FCoV transmits readily via the fecal-oral route, infects enterocytes in the gastrointestinal tract of cats, and causes only a mild infection. If the viral infection worsens and becomes systemic and lethal, then the virus is classified as the FIPV biotype. Symptoms associated with the FIPV biotype infection are granulomatous lesions, vascular leakage, and/or pleural and peritoneal effusion. The current understanding is that FIPV arises during *in vivo* infection due to genetic mutations of FECV (13, 23, 26, 36). In approximately 1-5% of enteric infections, such spontaneous "internal mutation(s)" extend FCoV tropism to include blood monocytes and tissue macrophages (20). The resulting productive infection of these cells, a hallmark of FIP, enables systemic spread and results in macrophage activation, with concomitant immune-mediated events leading ultimately to death. To date, the precise mutation(s) that account for a shift in the FCoV biotype have not been identified and there is no definitive diagnostic test for FIP, except through post-mortem histopathological analysis by immunocytochemistry.

In addition to the two different biotypes (FECV and FIPV), two FCoV serotypes have been identified. Serotype 1 FCoVs are highly prevalent clinically (3, 14, 33), but do not propagate in cell culture and are therefore studied less often than serotype 2 FCoVs, which are easily propagated *in vitro* but not as clinically prevalent. In this study, we chose to study FECV and FIPV from serotype I FCoV.

The FCoV genome is approximately 29 kB with 11 open reading frames encoding replication, structural, and accessory proteins (15). Like other RNA viruses, coronavirus replication is error prone, with an estimated mutation rate of approximately $4\times10^{-4}$ nucleotide substitutions/site/year (31, 37). It has been suggested that mutations in the spike protein (S), 3c, and/or 7b genes are associated with the transition of FECV to FIPV (1, 5-7, 19, 22, 24, 28). Since the FCoV spike protein mediates receptor binding (through the S1 subunit) and fusion (through the S2 subunit), and mutations in the S gene allowed FCoV to infect macrophages (30), we focused on correlating mutations in this S gene with the biotype switch and changes in cellular tropism.

The coronavirus spike protein (S) is a class I fusion protein, which typically requires activation by cellular proteases to be able to mediate cell entry. Mutation of the proteolytic cleavage sites can have profound implications for disease progression (4, 16), and therefore we sought to determine whether mutations at protease cleavage sites are evident in the FIPV biotype.

Proteases cleave their substrates though recognition of specific amino acid motifs within the relative position designation P6-…-P1 | P1'-…-P6', where cleavage occurs at the P1 position, adjacent to the P1' position (27). In this paper, we will denote the cleavage site between P1 and P1' using the vertical bar "|" symbol. For a typical serine protease like trypsin, there is a strong requirement for a basic amino acid (Arg or Lys) at the P1 cleavage position. Substrate residues flanking P1 can also have major impacts on the rate and specificity of cleavage for a given protease; in the case of influenza, the addition of arginine/lysine residues upstream of the P1 residue (the polybasic region) of the viral HA can allow an alternate protease called furin to cleave HA and increase the virulence of this virus (25).

Until recently, FCoVs were thought to use uncleaved S proteins to enter cells. However, a functional furin cleavage site has been identified in two examples of serotype 1 FECVs, located at the boundary of the S1 and S2 subunits (9). Furin is a ubiquitous proprotein convertase (PC) enriched in the trans-Golgi network and is well conserved among mammals (32). Furin cleaves a wide range of protein precursors into biologically active products at a consensus motif that is often defined as R-X-K/R-R, where R is the basic arginine residue, X is any residue, and K is the basic lysine residue (35). The minimum requirement for furin cleavage is for a P1 and P4 arginine. In addition to the P1-P4 arginines, these residues are often flanked by serine residues, and promoted by the presence of a basic residue at P2. The general cleavage requirement for a PC enzyme is defined as cleavage at paired basic residues (B-$X_{2n}$-B), where the intermediate residues (-$X_{2n}$-) consist of 0, 2, or 4 amino acids (32).

For many enveloped viruses, protease cleavage occurs adjacent to the fusion peptide site, which is located at the boundary of the receptor-binding and fusion domains of the envelope protein. For coronaviruses the fusion peptide is located within the fusion domains, in close proximity to second processing site denoted as S2'. The actual proteases that cleave at this site are currently unknown, but may include members of the cathepsin family that function in endosomes and lysosomes. Mutations at the S2' cleavage site could be important and therefore we have chosen to study this site in addition to the S1/S2 cleavage site. In this work, we used an approach to studying FIP that complements and extends previous work. We focused on sequence analysis of the two cleavage-activation sites of the spike gene (S1/S2 and S2'), which are functionally relevant regions for virus entry and infection. To obtain FECV gene samples, fecal material from healthy cats that carried FCoV was sequenced for the S gene. These gene samples were operationally defined as being from the FECV biotype, and 56 FECV samples were obtained in this study. To obtain FIPV samples, tissue and body fluid samples were collected from cats that were diagnosed with FIP. For tissue samples, cats were typically euthanized based on a clinical diagnosis of FIP, and samples were collected post-euthanasia. In many cases, a diagnosis of FIP was reconfirmed by immunohistochemistry (IHC) of tissue from the euthanized cats, but not all samples were examined by IHC for logistical reasons. Viral gene sequences obtained from FIP cats were operationally defined as being from the FIPV biotype. In this study, 84 FIPV samples were sequenced from 39 cats and various organs and tissues. Samples from our previous study (19) were also included to provide as complete a data set as possible, and we compared our data to those of a separate study by Chang *et al.* (6). Our studies were consistent with the hypothesis that mutations in both the S1/S2 and S2' protease cleavage sites were present in FIPV samples.

## Methods

*FCoV sample acquisition and sequence analysis.* Fecal samples from subclinically infected cats and tissue samples from cats with a clinical diagnosis of FIP were solicited from shelters, breeders and veterinarians throughout the United States. To initially confirm samples were FCoV-positive, RNA was extracted using QIAamp Viral RNA Mini (Qiagen, Valencia, CA) and FCoV primers that detect most circulating strains were used to screen all fecal samples using the procedure outlined by Herrewegh and coworkers (12). RNA extracted from the FIPV-TN406/Black lab-adapted strain was used as positive control. Fresh tissue samples from FIP-diagnosed cats were harvested and RNA samples were extracted using MagMAX Express (Life Technologies, Grand Island, NY).

For all samples that were FCoV-positive, 50 μL RT-PCR reactions were performed with One-Step RT-PCR (Qiagen, Valencia, CA) using gene-specific S primers, encompassing S1/S2 and S2': S1/S2 primer pair

*5'*-GCACAAGCAGCTGTGATTA-*3'* and *5'*-GTAATAGAATTGTGGCAT-*3'*; S2' primer pair *5'*-GATATGATCACAGTATCAGATCG *-3'* and *5'*-ATAATC ATCATCAACAGTGCC-*3'.* PCR conditions were: 30 min at 50°C, 15 min at 95°C, and 39 or 35 cycles of: 1 min at 94°C, 1 min at 55°C, 1 or 1.5 min at 72°C, 10 min at 72°C. PCR products were purified using Qiaquick Gel Extraction (Qiagen, Valencia, CA). Sanger-based dideoxy sequencing was performed at the Life Sciences Core Laboratories (Cornell University). DNA sequences were translated into protein sequences and alignments were performed using Geneious 5.4 (http://www.geneious.com).

*Visual Representation of Mutation Data.* To visualize the biochemical changes in the protease cleavage sites of the S protein, we developed a specialized scatter plot through MATLAB [MathWorks, Natick, MA]. A separate plot was made for each residue position being evaluated, and the Van der Waals (VdW) volume of amino acids (29) was plotted against their hydropathy index (HPI) (18). In general, amino acids with lower HPI are more hydrophilic whereas those with higher HPI are more hydrophobic. Each data point corresponds to an amino acid that was found at the residue position being evaluated, and the size of the data point scales with the frequency of occurrence of that amino acid within a data set. We caution here that the data point size does not denote an error range in VdW volume or HPI. Other important properties of the amino acids were portrayed by color-coding the data points as such: polar positive (blue: H,K,R), polar negative (red: D,E), polar neutral (green: S,T,N,Q), nonpolar aliphatic (gray: A,V,L,I,M), nonpolar aromatic (magenta: F,W,Y), unique (brown: P,G), and disulfide bond-forming (cyan: C).

## Results and Discussion

### FECV sequence from healthy cats

The amino acid sequence motifs for the S1/S2 and S2' regions are shown in Table 1. For the S1/S2 site, the previously recognized motif -R-R-S/A-R-R-S- (19) was again found to be highly conserved. There was a limited variability in the P5 position, with Arg substituted for: Gly (1 cat) and Lys (1 cat). The functional relevance of this variation is unclear, though the P5 position is not generally considered to be highly important for furin cleavage. There was also a single case with a change in the P3 position with Ser/Ala substituted for Val (1 cat). As with P5, the type of amino acid residue at the P3 position is not functionally critical for furin cleavage. Based on the expanded dataset available from this study, we also noted a high degree of conservation at the P6 cleavage position (which can be highly important for furin cleavage). All samples contained a polar uncharged residue (S, T or Q) at the P6 position. Overall, the expanded data set reinforces the notion that the S1/S2 motif is cleaved by furin. Based on the data from this study, we consider to following as an "FECV" motif at the S1/S2 position: -S/T/Q-R-R-S/A-R-R-S-

At the S2' site, the conserved nature of the FCoV fusion peptide (SAVEDLLF) was readily apparent, along with the expected conserved arginine residue in the P1 cleavage position (Table 1). All FECV samples tested contained a P2 lysine, with the high level of conservation implicating this as a functionally critical residue. The P1' residue (S) was also 100% conserved, consistent with a role in protease cleavage, as well as in fusion peptide function. Overall, we found an invariant motif -K-R-S- at the S2' position of FECV, and the dibasic nature of this site is consistent with cleavage by a range of proteases, including non-furin PCs and cathepsin B.

### FIPV sequence from FIP cats

The amino acid sequence motifs for the S1/S2 and S2' regions in the FCoV spike gene amplified from various tissues and body fluid samples are shown in Table 2. For S1/S2, the FECV motif -S/T/Q-R-R-S/A-R-R-S- was disrupted in the vast majority of samples. In addition, the S2' FECV -K-R-S- motif was disrupted in many samples. Fecal material from FIP cats is shown in Table 3, with sequences very similar to FECV samples.

For S1/S2, we focused our analysis on the "core" residues for furin cleavage, i.e. the P1', P1, P2 and P4 residues. The most frequent substitution was at the P1 position, where 11 cats displayed a change of Arg for Gly (4 cats), Met (4 cats), Trp (2 cats), Thr (1 cat) and Ser (1 cat). The next most frequent substitution was at the P2 position, where 10 cats displayed a change of Arg for Leu (6 cats), Pro (3 cats), Ser (1 cat) and His (1 cat). Five cats had a substitution at the P4 position, which displayed a change of Arg for Thr (2 cats) Gly (1 cat), Ser (1 cat), and Met (1 cat). Four cats had a substitution at the P1' position, which displayed a change of Ser for Ala (2 cats) Pro (1 cat), Gly (1 cat), and Leu (1 cat). All of these changes in the core residues would be expected to be highly functionally significant for furin cleavage.

At the S2', the conserved nature of the fusion peptide (SAVEDLLF) was again readily apparent, along with an invariant Ser at the P1' position (Table 2). While most FIPV cats contained a P1 arginine, three cats had substitutions for Ser. This change would be expected to be highly functionally significant for protease cleavage. There were more substitutions in the P2 residue, with five cats displaying a change of Lys for Met (3 cats) and Glu (2 cats). These changes would also be expected to be highly functionally significant for protease cleavage. The overall switch in the dibasic nature of the S2' site in some cats is consistent with a change in the protease(s) cleaving at this position.

For the core residues at S1/S2 and S2'. 31 out of 39 FIP cats (77%) had samples that were modified from the FECV consensus (-S/T/Q--R-R-S/A-R-R-S-). Healthy cats showed no deviation from the core motif

The P6, P5 and P3 residues within the S1/S2 site (-S/T/Q--R-R-S/A-R-R-S-) were defined as "non-core" residues as these are likely to be less functionally relevant for furin cleavage. Of these residues, P6 likely impacts furin cleavage the most. For the P6 position, five cats displayed a change of Ser/Thr/Gln for Pro (2 cats), Phe (1 cats), Ala (1 cat) and Leu (1 cat). All of these changes would be expected to be functionally significant for furin cleavage.

Including S1/S2 P6 as a core residue revealed that 33 out of 39 FIP cats (85%) had samples that were modified from the FECV consensus (-S/T/Q-R-R-S/A-R-R-S-). Healthy cats showed no deviation from a core motif including P6.

For the non-core S1/S2 P5 position, nine cats displayed a change of Arg for Lys (8 cats), as seen for FECV. For the non-core S1/S2 P3 position, two cats displayed a change of Ser/Ala for Met (1 cats) and Thr (1 cat). The significance of these changes is uncertain. The S2' non-core P1' residue showed no deviation in either FIP or healthy cats.

Including both core and non-core residues at S1/S2 and S2', we found that 37/39 FIP cats (95%) had mutations compared to healthy cats. 9% of healthy cats showed a mutation within the non-core residues

It is noteworthy that many cats had multiple changes across the two proteolytic cleavage sites and while there was a wide variety of different substitutions present in different cats, sampling of multiple tissues within an individual cat (e.g. cat#197) revealed relatively limited variation. It is also noteworthy that most of our samples would be expected to be comprised of "end-stage" viruses, where cats have more obvious clinical signs of FIP. It will be interesting to track cats that might be in an "early", possibly sub-clinical, phase of FIP. These cats may harbor viruses that have different mutations to end-stage viruses sampled here, in particular in the "non-core" residues.

While a prediction of the proteases coming into play for FIPV is quite uncertain, the trend in our data is for the replacement of critical basic residues and serine with more hydrophobic residues, a pattern consistent with cleavage by an alternative protease(s). Such proteases may include cathepsins (8), or matrix metalloproteases (MMPs) (17). MMPs are known to be highly expressed on macrophages (2) and to our knowledge, MMP-mediated cleavage of a viral envelope protein would be unique situation for a virus entry pathway. Both MMPs and cathepsin demonstrate relatively broad substrate specificity, in line with the range of mutations seen in the cleavage site(s) of FIPVs.

**Comparison to a separate study (Chang *et. al.* 2012) and to database sequences**

The data generated during the course of our study was compared to that generated by a separate study, Chang *et al.* 2012 (6), as well as to additional database sequences (Table 4). For S1/S2, of the ten FECV sequences in Chang *et al.*, the -Q/R/S-R-R-S/A-R-R-S- motif was preserved with one exception, where one cat had a P6 Pro residue. All core residues were identical to our study. A P6 Pro was also found in one other database sequence. The pattern of residues in FIPV samples followed the same general pattern as described for our study; i.e. introduction of non-polar, aliphatic or aromatic residues in the protease recognition sites. Of the eleven FIP cats samples in Chang *et al.* 2012, four had P1 substitutions (G, W and S), two had P1' substitutions (L and P), two had P4 substitutions (G and S), and three had P6 substitutions (L and P). Of seven database samples, three had P1 substitutions (G, W and S), two had P1' substitutions (P), one had P4 substitutions (Q), and five had P6 substitutions (A and P).

At S2', Chang *et al.* 2012 show some limited changes of the invariant KRS motif for FECV, with different P2 residues in four cats, substituted with Arg (3 cats) and Thr (1 cat). Cats with FIP showed an introduction of a hydrophobic reside at P2 in five cases, with Lys being substituted for Met (4 cats) and Val (1 cat). A similar trend was present in database samples, with introduction of Val, Ala and Gln in three individual cases.

Overall, we consider that the data from Chang *et al.* 2012, as well as other database sequences, are in strong agreement with the data from this study. 10/11 (91%) of samples from Chang *et al* 2012*.* and 4/4 (100%) of database samples had functionally significant mutations in the spike cleavage sites.

### Visual representation of sequences in the protease activation sites of FECV and FIPV spike proteins

Finding a mutation pattern that could explain the transformation of FECV to FIPV is challenging because amino acids have multiple properties that are similar or dissimilar with each other, such as molecular size, hydrophobicity, and charge (34). In order to simultaneously visualize changes to multiple biochemical properties, we developed a specialized scatter plot through MATLAB that shows the Van der Waals volume (Y-axis), hydropathy index (X-axis), frequency of occurrence (data point size), and unique property (color code) of amino acids found at the cleavage site positions. The resulting plots are shown in Figure 1 for the residues at the S1/S2 and S2' cleavage sites for both the FECV and FIPV samples.

Compared to FECV, FIPV contains the following changes in the S1/S2 cleavage site: P6 and P1' mutates from polar neutral residues to nonpolar residues, P4 and P2 mutates from arginine to smaller and more hydrophobic residues, and P1 mutates from arginine to more hydrophobic residues. At the S2' cleavage site, FIPV contains the following mutations: P2 mutates from positively charged lysine to either negatively charged glutamate or uncharged methionine, and P1 mutates from positively charged arginine to polar neutral serine that is half the volume of arginine. In summary, the visualization of biochemical changes at the cleavage sites using our plotting method can greatly aid with the identification of mutation patterns associated with FIP.

### Potential use of this study for diagnosis of FIP in cats

We believe that our study will be of benefit in the diagnosis of FIP in cats by detecting mutations in S protein cleavage site that mimic those found in FIPV. However, the pattern of variant viruses seen in FIP cats is complex. We consider that the tracking of the ten amino acids that comprise the cleavage sites of FCoV spike (-S/T/Q-R-R-S/A-R-R-S- from S1/S2 and -K-R-S- from S2') with respect to the biochemical parameters of size, hydrophobicity, and charge can be used to predict the likelihood of FIP in cats. To simplify the detection of FIPV-like mutations, one could potentially use the changes in these biochemical parameters in conjunction with a ranking of the core and non-core residues to arrive at a "FIP diagnostic score." Such a scoring system could help clinicians identify which cats are likely to develop FIP without the need to functionally investigate every mutation found. The use of this scoring metric for samples that clinicians can easily access (e.g. blood) would be helpful for diagnosing FIP early and quickly. While our current study only utilized a limited number of blood samples, mutations consistent with a FIPV biotype were seen in all positive samples. However, it is known that FCoV can be present in the blood monocytes of healthy cats without necessarily leading to FIP (10, 12, 28). Further investigation into the spike protein cleavage site changes in blood samples of both healthy and FIP cats will form a future focus of this study.

We considered that fecal samples from FIP cats were likely to be contaminated with an ongoing infection with FECV, and so not discriminatory for FIP. This was confirmed, as the FIP cats tested in this study continued to shed FECV-like viruses, based on the highly limited sequence variation observed at the spike cleavage sites (Table 3).

FIP exists in two clinical manifestations (wet and dry) (21). As in our previous study (19), we found no correlation between sequence alterations at the spike cleavage sites between "wet" and "dry" FIP cats. These differences in clinical presentation are likely to be due to immunological factors in individual cats, rather than differences in the cleavage-activation of the viruses infecting these cats.

The use of the techniques reported here for serotype I FCoVs may also be applied to serotype II FCoVs; however, these viruses lack an equivalent S1/S2 motif. Investigation of predictive changes in the S2' region of serotype II FCoVs is currently ongoing.

**Acknowledgments**

We thank Jean Millet, Lisa Bolin and members of the Whittaker laboratory, Ruth Collins, Dante Lepore and members of Collins laboratory, and Susan Daniel and members of Daniel laboratory for advice and discussion during the course of this study. The authors also thank Drs. Meredith Brown, Stephen O'Brien, Sean McDonough, Edward Dubovi and Gerald Duhamel for providing some of the clinical samples used in this study, Andrew Regan and Kirsten Elfers for initial work on this project, and Nadia Chapman, Vera Rinaldi, Misty Pocwiercz, Valerie Marcano, Rod Getchell and Wendy Wingate for technical assistance. This work was funded in part by a sponsored project from Antech Diagnostics. B.L. was supported by funds from the Cornell University College of Veterinary Medicine DVM/PhD program. Work in the author's lab is also supported by grants from the Cornell Feline Health Center, the Winn Feline Health Foundation and the Morris Animal Foundation. The sponsors had no influence in the study design, the collection, analysis and interpretation of data, the writing of the manuscript, or in the decision to submit the manuscript for publication. All work with animals was approved by the Institutional Animal Use and Care Committee at Cornell University (Ithaca NY).

**TABLE 1**

Amino acid sequences in the S1/S2 and S2' regions of the FCoV spike obtained from fecal samples from heathly cats. The ten amino acids comprising the predicted proteolytic cleavage sites are in bold, with modifications to these residues colored (green = modification to non-core residue).

| Cat ID | Sample | S1/S2 | S2' |
|---|---|---|---|
| 150 | Feces | TS**SRRSRRS**TTE | HSIG**KRS**AVED |
| 106 | Feces | TQ**SRRSRRS**YPD | PTIG**KRS**AVED |
| 110 | Feces | TQ**TRRSRRS**TSE | PRIG**KRS**AVED |
| 111 | Feces | TH**SRRARRS**TVE | PRIG**KRS**AVED |
| 125 | Feces | TQ**SRRARRS**TVE | PTIG**KRS**AVED |
| 126 | Feces | TQ**SRRSRRS**ASS | PTIG**KRS**AVED |
| 128 | Feces | TH**SRRARRS**TVE | PTIG**KRS**AVED |
| 129 | Feces | TQ**SRRSRRS**TSD | PTIG**KRS**AVED |
| 131 | Feces | TQ**SRRSRRS**ASN | PTIG**KRS**AVED |
| 132 | Feces | TQ**SRRSRRS**APE | PRIG**KRS**AVED |
| 135 | Feces | TQ**SRRARRS**LPA | PRIG**KRS**AVED |
| 136 | Feces | TQ**SRRSRRS**VVE | PQIG**KRS**AVED |
| 137 | Feces | TS**SRRSRRS**TPE | HSIG**KRS**AVED |
| 138 | Feces | TQ**SRRSRRS**VAE | PTIG**KRS**AVED |
| 140 | Feces | TQ**SRRSRRS**VVE | PTIG**KRS**AVED |
| 141 | Feces | TQ**SRRSRRS**VVE | PQIG**KRS**AVED |
| 142 | Feces | TH**SRRARRS**TVE | PTIG**KRS**AVED |
| 143 | Feces | TS**SRRARRS**SVE | PTIG**KRS**AVED |
| 144 | Feces | TQ**SRRSRRS**ASM | PTIG**KRS**AVED |
| 146 | Feces | TH**SRRARRS**TVE | PKIG**KRS**AVED |
| 149 | Feces | TS**SRRSRRS**TPE | HSIG**KRS**AVED |
| 150 | Feces | VNHTS**SRRSRRS**TTET | HSIG**KRS**AVED |
| 152 | Feces | TH**SRRSRRS**NSD | PRIG**KRS**AVED |
| 153 | Feces | PH**SRRSRRS**TNY | PTIG**KRS**AVED |
| 155 | Feces | TQ**SGRSRRS**ASD | not determined |
| 160 | Feces | NHTH**SKRSRRS**TSN | PTIG**KRS**AVED |
| 167 | Feces | TH**SKRSRRS**TSN | PTIG**KRS**AVED |
| 234.1 | Feces | TH**TRRSRRS**APV | PTIG**KRS**AVED |
| 246 | Feces | NHTQ**SKRSRRS**APH | PKIG**KRS**AVED |
| 304 | Feces | TH**TRRSRRS**APV | PTIG**KRS**AVED |
| 307 | Feces | TH**TRRSRRS**APV | PTIG**KRS**AVED |
| 308 | Feces | NHTH**TRRSRRS**API | PKIG**KRS**AVED |
| 308 | Feces | TH**TRRSRRS**API | PTIG**KRS**AVED |
| 310 | Feces | TH**TRRSRRS**APA | PTIG**KRS**AVED |
| 313 | Feces | HAR**TRRSRRS**APV | PKIG**KRS**AVED |
| 349 | Feces | NTQ**SRRARRS**ASDS | PQIG**KRS**AVED |
| 352 | Feces | NTQ**SRRARRS**ASDS | PQIG**KRS**AVED |
| 277.2 | Feces | NHTQ**SRRSRRS**TSDFV | PTIG**KRS**AVED |
| 278 | Feces | **NHTQSRRVRRS**VQESVQ | PTIG**KRS**AIED |
| BL1 | Feces | TH**TRRSRRS**APA | PTIG**KRS**AVED |
| BL10 | Feces | TH**TRRSRRS**APV | PTIG**KRS**AVED |
| BL11 | Feces | TH**TRRSRRS**APV | PTIG**KRS**AVED |
| BL12 | Feces | TH**TRRSRRS**APV | PTIG**KRS**AVED |
| BL13 | Feces | TH**TRRSRRS**APV | PTIG**KRS**AVED |
| BL2 | Feces | TH**TRRSRRS**APA | PTIG**KRS**AVED |
| BL3 | Feces | TH**TRRSRRS**APA | PTIG**KRS**AVED |
| BL4 | Feces | TH**TRRSRRS**APA | PTIG**KRS**AVED |
| BL5 | Feces | TH**TRRSRRS**APV | PTIG**KR**SAVED |
| BL6 | Feces | TH**TRRSRRS**API | PTIG**KRS**AVED |
| BL7 | Feces | TH**TRRSRRS**API | PTIG**KRS**AVED |
| BL8 | Feces | TH**TRRSRRS**API | PTIG**KRS**AVED |
| BL9 | Feces | TH**TRRSRRS**APA | PTIG**KRS**AVED |
| FECV-4582 | Feces | TQ**QRRSRRS**TSD | PTIG**KRS**AVED |
| FECV-4594 | Feces | TQ**QRRSRRS**TSD | PTIG**KRS**AVED |
| FECV-FCA4597 | Feces | TQ**QRRSRRS**TSD | PTIG**KRS**AVED |
| FECV-FCA4606 | Feces | TQ**QRRSRRS**TSD | PTIG**KRS**AVED |

**TABLE 2**
Amino acid sequences in the S1/S2 and S2' regions of the FCoV spike obtained from tissue and body fluid samples from cats diagnosed with FIP. The ten amino acids comprising the predicted proteolytic cleavage sites are in bold, with modification to these residues colored (red and blue = modification to core residue, green and violet = modification to non-core residue). ND = not determined.

| Cat ID | Tissue | S1/S2 | S2' |
|---|---|---|---|
| 129308 | Mesentery | TS**SRRSPRS**TLD | not determined |
| | Lower Gut | TS**SRRSLRS**TVR | not determined |
| 147 | Blood | TQ**SRRARSS**ASD | PRIG**ERS**AVED |
| 148 | Ascites | TS**SRRSRRP**TTE | HSIG**KRS**AVED |
| 151643-08 | Heart | TQ**FRRARRS**AVR | not determined |
| | Spleen | TQ**FRRSRRS**TPG | not determined |
| | Liver | TQ**FRRSRRS**TVR | not determined |
| 157 | Ascites | NHTQ**PRRARRS**VSELV | PPSIG**KRS**AVED |
| | Liver | NHTQ**PRRARRS**VSELV | PPSIG**KRS**AVEDLLFN |
| | Spleen | NHTQ**PRRARRS**VSELV | PSIG**KRS**AVEDLLF |
| | Spleen | NHTQ**PRRARRS**VSELV | PPSIG**KRS**AVEDLLFN |
| 159 | Ascites | NHTS**SRRSRMS**TPET | PPSIG**KRS**AVED |
| 162 | Ascites | NHTQ**SRRSRRA**TSNP | PTTG**KRS**AVED |
| 163 | Urinary Discharge | NHTQ**SRRSRMS**TSDS | PPRVG**KSS**AIED |
| 166 | Brain | HTH**PRRSRGS**TIETV | not determined |
| 170 | Ascites | TQ**SRRSRRS**TSD | PTIG**KRS**AVED |
| | Liver | not determined | PTIG**KRS**AVED |
| | Lung | NHTQ**SRRSRRS**TSDFVT | PPTIG**KRS**AVEDLLFN |
| 179 | Ascites | NHTQ**SRGSRRS**TSD | PTIG**KRS**AVED |
| 181 | Chest Fluid | NHTQ**SRRSRRS**TSD | PPTIG**KRS**AVED |
| | Liver | VNHTQ**SRRSRRS**TSD | PTIG**KRS**AVEDLLFN |
| | Lung | VNHTQ**SRRSRRS**TSD | PTIG**KRS**AVEDLLFN |
| | Spleen | VNHTQ**SRRSRRS**TSDF | PTIG**KRS**AVEDLLFN |
| 182 | Ascites | NHTQ**SRSSRRS**TPV | PPSIG**KRS**AVED |
| 192 | Thoracic Fluid | NHTQ**SRRSRGS**TSD | PPTIG**KRS**AVED |
| 195 | Spinal Cord | NHTQ**SKTAPRG**TSDSV | not determined |
| | | NHTQ**SKTAPRA**TSDS | not determined |
| 197 | Ascites | NHTH**SKRARWS**TSD | PRIG**MRS**AIED |
| | Kidney | TH**SRRSLRS**APV | not determined |
| | Liver | NHTH**SKRARWS**TSDS | PPRIG**MRS**AIEDLLFN |
| | Lung.1 | TH**TRRSLRS**APD | not determined |
| | Lung.2 | NHTH**SKRARWS**TSDS | PPRIG**MRS**AIEDLLFN |
| | Omentum | NHTH**SKRARWS**TSDS | PPRIG**MRS**AIEDLLFN |
| | Small Intestine.1 | TH**SRRSRRS**ASD | not determined |
| | Small Intestine.2 | NHTH**SKRARWS**TSDS | PPRIG**MRS**AIEDLLFN |
| | Spleen | NHTH**SKRARWS**TSDS | PPRIG**MRS**AIEDLLFN |
| 211 | Ascites | NHTH**LRRSRRS**TPE | RVG**KRS**AVED |
| 215 | Ascites | NHTQ**SKSARRS**TSD | PRIG**KRS**AIED |
| 232 | Liver | TQ**SKSARRS**TSD | not determined |
| | Lung | TQ**SKSARRS**TSD | not determined |
| | Spinal Fluid | NHTQ**SKSARRS**TSD | PRIG**KRS**AIED |
| 234 | Kidney | TH**TRRSLRS**APV | not determined |
| | Lung | NHTH**TRRSLRS**APVA | PPTIG**KRS**AVEDLLF |
| 239 | Blood | NHTQ**SKTARRS**TSD | PRVG**KSS**AIED |
| | Brain | NHTQ**SKTARRS**TSDS | PPRVG**KSS**AIED |
| | | NHTQ**SKTALRS**TSDS | not determined |
| | Cerebellum | NHTQ**SKTALRS**TSDSVI | PPRVG**KSS**AIED |
| | | NHTQ**SKTARRS**TSDSVI | not determined |
| | Cerebrum | NHTQ**SKTARRS**TSDS | PPRVG**KSS**AIED |
| | Kidney.1<br>Kidney.2<br>Spleen | THTRRSLRSAPV<br>NHTQSKTARRSTSDS<br>NHTQSKTARRSTSD | not determined<br>PPRVGKSSAIED<br>PPRVGKSSAIEDLLF |
| 241 | Ascites | NYTHSRRSLSSTPE | PRIGKRSAVED |
| 244 | Mesentery<br>Mesenteric lymph node | HSRRASTSTSN<br>TQSRRASTSTSN | not determined<br>not determined |
| 245 | Spinal Fluid | NHTQSRRSRRSKSN<br>not determined | PTVGKSSAVED<br>TVGKRSAVED |
| 256 | Ascites | NHTHSKRARWSTSD | PRIGMRSAIED |
| 259 | Brain<br>Lung | THTQSKSARRSTSDSVI<br>TQSKSARRSTSD | KRSAIED<br>not determined |
| 261 | Liver<br>Small Intestine<br>Spleen | TQSRRMRRSTSN<br>TRSRRMRRSTSN<br>TQSRRMRRSTSN | not determined<br>not determined<br>not determined |
| 350 | Kidney<br>Lung | NHTSARRSRRSASEII<br>VNHTSARRSRRSASEI | PTIGKRSAVED<br>PTIGKRSAVED |
| 77 | Kidney<br>Cerebellum.1<br>Cerebellum.2 | THSRRSRMSTQN<br>THSRRSLRSTQN<br>THSRRSRRSTQN | not determined<br>not determined<br>not determined |
| 08-153990 | Kidney<br>Cerebrum<br>Brain stem<br>Cerebellum | TQPRRARMSVPE<br>TQPRRARMSVPE<br>TQPRRAPMSVPE<br>TQPRRARMSVPE | not determined<br>PTIGMRSAVED<br>PTIGMRSAVED<br>PTIGMRSAVED |
| N04-93 | Kidney<br>Cerebrum | THLRRSHRSTSE<br>TLTGRSHRSTSE | not determined<br>not determined |
| N05-48 | Cerebellum | TQSRSSRRSTSD | not determined |
| N05-110 | Mesenteric lymph node<br>Cerebellum | TQTKRSRRSTPQ<br>TQTKRSRRSTPA | PTIGKRSAVED<br>not determined |
| N07-95 | Cerebrum | THTRKTRRSIAD | not determined |
| D04-397 | Spleen<br>Mesentric lymph node | TQSRRSRRSTVD<br>TQSRRSRRLTSN | not determined<br>not determined |
| D06-327 | Spleen<br>Mesentery | THSRRSRGSAPN<br>THSRRSRGSAPN | not determined<br>not determined |
| 429 | Omentum | TSSRSSRRSTSE | SRIGERSAVED |
| 430 | Omentum | SQSRRSRSSTSE | PRVGKRSAVED |

**TABLE 3**
Amino acid sequences in the S1/S2 and S2' regions of the FCoV spike obtained from fecal samples from cats diagnosed with FIP. The ten animo acids comprising the predicted proteolytic cleavage sites are in bold, with modification to these residues colored (green = modification to non-core residue).

| Cat ID | Sample | S1/S2 | S2' |
|---|---|---|---|
| 107 | Feces | TT**SRRPRRS**DPA | PTIG**KRS**AVED |
| 108 | Feces | QS**SRRSRRS**TSD | PTIG**KRS**AVED |
| 145 | Feces | TH**SRRARRS**TVE | PKIG**KRS**AVED |
| 148 | Feces | TS**SRRSRRS**TTE | PKIG**KRS**AVED |
| 215 | Feces | TH**ARRSRRS**TPE | PRIG**KRS**AIED |
| 251 | Feces | TQ**SKRARRS**TSD | PRIG**KRS**AIED |
| 347 | Feces | TR**SRRSRRS**TLEP | PRVG**KRS**AVED |
| 352 | Feces | TQ**SRRARRS**ASDS | PQIG**KRS**AVED |

**TABLE 4**

Amino acid sequences in the S1/S2 and S2' regions of the FCoV spike obtained from Chang et al. and databases. The ten amino acids comprising the predicted proteolytic cleavage sites are in bold, with modification to these residues colored (red and blue = modifiation to core residue, green = modification to non-core residue). Modifications that are not expected to be functionally relevent are shown in gray. n/a - not applicable/unknown.

| **FECV** | | | |
|---|---|---|---|
| **Name** | **Tissue** | **S1/S2** | **S2′** |
| UU2 | feces | **SRRSRRS** | **RRS** |
| UU7 | feces | **SRRARRS** | **KRS** |
| UU10 | feces | **SKRSRRS** | **KRS** |
| UU11 | feces | **SKRSRRS** | **KRS** |
| UU18 | feces | **PRRSRRS** | **KRS** |
| UU19 | feces | **SRRSRRS** | **TRS** |
| UU20 | feces | **SRRSRRS** | **KRS** |
| UU22 | feces | **SRRSRRS** | **RRS** |
| UU23 | feces | **SRRSRRS** | **RRS** |
| UU31 | feces | **SRRSRRS** | **KRS** |
| DATABASE SEQUENCE | | | |
| RM | n/a | **PRRSRRS** | **KRS** |

| **FIPV** | | | |
|---|---|---|---|
| **Name** | **Tissue** | **S1/S2** | **S2′** |
| UU3 | n/a | **SRRSRRS** | **MRS** |
| UU4 | n/a | **SRSARGS** | **MRS** |
| UU5 | n/a | **SRRSRGS** | **KRS** |
| UU8 | n/a | **SRRSRRS** | **KRS** |
| UU9 | n/a | **SRRSRRL** | **KRS** |
| UU15 | n/a | **LRRSRRP** | **KRS** |
| UU16 | n/a | **PRGSRRS** | **MRS** |
| UU17 | n/a | **SRRSRRS** | **VRS** |
| UU21 | n/a | **SRRSRWS** | **MRS** |
| UU24 | n/a | **SRRSRSS** | **KRS** |
| UU30 | n/a | **LRRSRRS** | **KRS** |

DATABASE SEQUENCES

| | | | |
|---|---|---|---|
| Black/TN406 | Liver | AKRSRRP | VRS |
| FCoV C1Je | Jejunum | PRQSRRS | KRS |
| Q66951_9ALPC (KU-2) | n/a | ARRSRSS | KRS |
| Q8JVL1_9ALPC (Black/TN406) | n/a | AKRSRRP | ARS |
| Q8JVL2_9ALPC (UCD1) | n/a | SRRSRGS | QRS |

**FIGURE 1.**

Visual representation of amino acid properties in the S1/S2 and S2’ cleavage sites for FECV and FIPV. The Y-axis is the Van der Waals volume (Å3), and the X-axis is the hydropathy index (unitless). The data point size corresponds to the frequency of occurrence. Note that for FIPV, the frequency of occurrence is based on all samples collected and not just the number of FIP cats sampled. The table/legend summarizes all amino acid properties and color code.

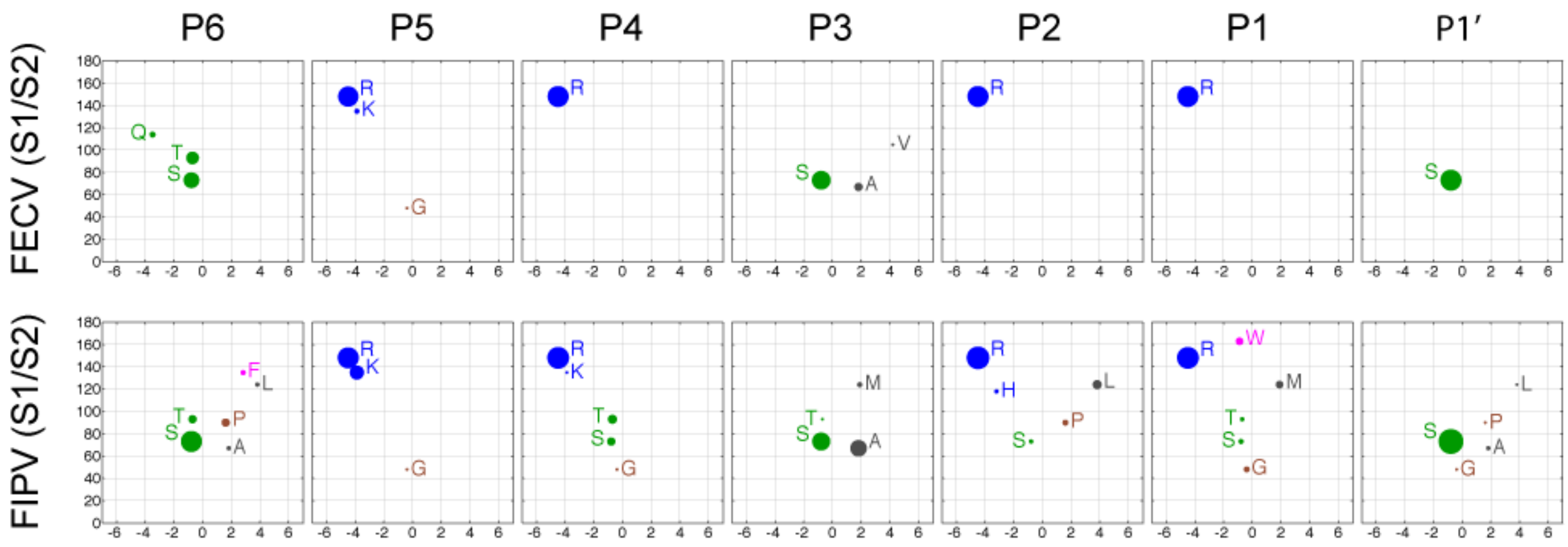

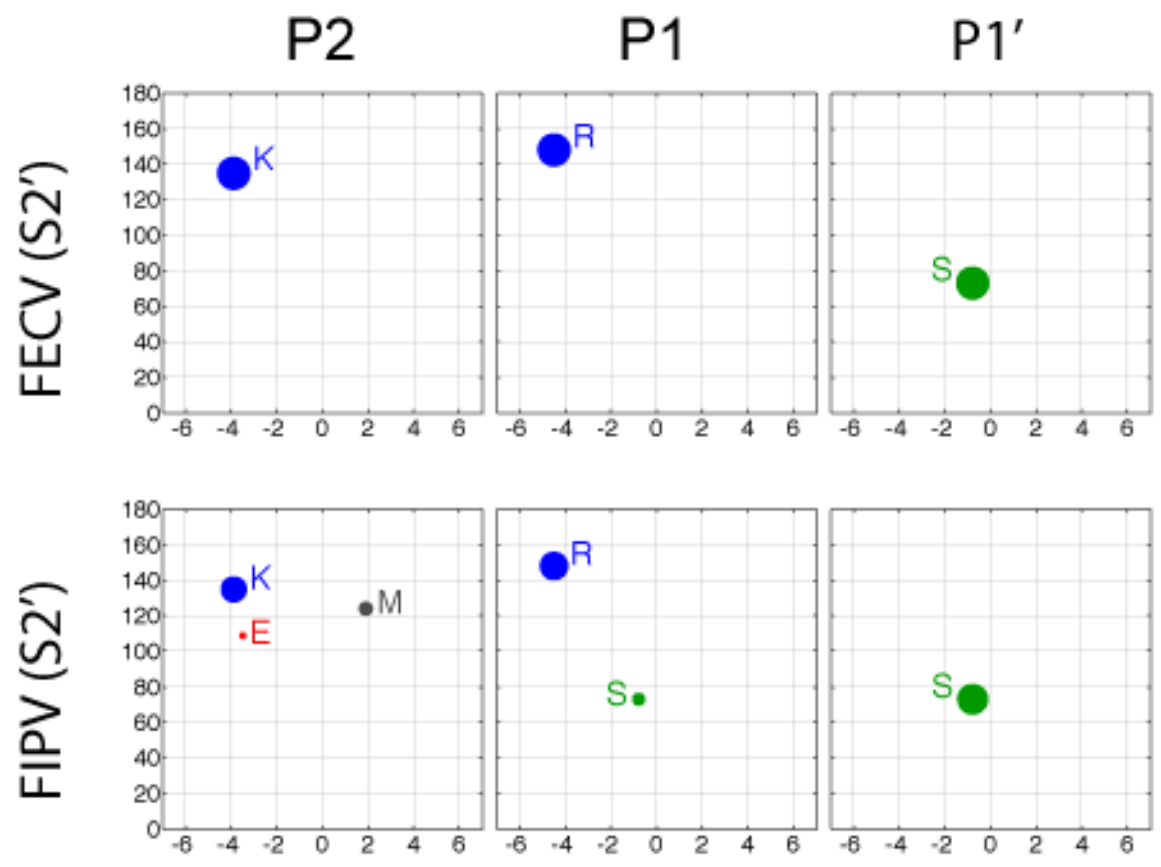

| Color | Property | Residue | Hydropathy Index | Van der Waals Vol. (Å$^3$) |
|---|---|---|---|---|
| | Polar Positive | K | -3.9 | 135 |
| | | H | -3.2 | 118 |
| | | R | -4.5 | 148 |
| | Polar Negative | D | -3.5 | 91 |
| | | E | -3.5 | 109 |
| | Polar Neutral | S | -0.8 | 73 |
| | | T | -0.7 | 93 |
| | | N | -3.5 | 96 |
| | | Q | -3.5 | 114 |
| | Nonpolar Aliphatic | A | 1.8 | 67 |
| | | V | 4.2 | 105 |
| | | L | 3.8 | 124 |
| | | I | 4.5 | 124 |
| | | M | 1.9 | 124 |
| | Nonpolar Aromatic | F | 2.8 | 135 |
| | | Y | -1.3 | 141 |
| | | W | -0.9 | 163 |
| | Unique | G | -0.4 | 48 |
| | | P | 1.6 | 90 |
| | Disulfide Bond Forming | C | 2.5 | 86 |